\documentclass[11pt]{article}
\usepackage{hyperref}
\usepackage[margin=1.2in]{geometry}
\usepackage{slantsc}
\pdfoutput=1
\begin{document}
\title{ \LARGE Modeling air entrainment in plunging jet using 3DYNAFS\textsuperscript{\textcopyright}}
\author{Chao-Tsung Hsiao, Jingsen Ma, Xiongjun Wu and Georges L. Chahine \\
DYNAFLOW, INC. 
Jessup, MD, 20794, USA
\\ Email: georges@dynaflow-inc.com
\\Tel: (301)604-3688}

\date{}

\maketitle

As the liquid jet plunges into a free surface, significant air is entrained into the water and forms air pockets. These air pockets eventually break up into small bubbles, which travel downstream to form a bubbly wake. To better understand the underlying flow physics involved in the bubble entrainment, in the linked videos, air entrainment due to a water jet plunging onto a pool of stationary water was numerically studied by using the 3DYNAFS\textsuperscript{\textcopyright}  software suit.

The flow field is simulated by directly solving the Navier-Stokes equations through the viscous module, 3DYNAFS-VIS\textsuperscript{\textcopyright}, using a level set method for capturing the free surface. The breakup of entrained air pockets and the resulting bubbly flow were modeled by coupling 3DYNAFS-VIS\textsuperscript{\textcopyright} with a Lagrangian multi-bubble tracking model, 3DYNAFS-DSM\textsuperscript{\textcopyright} \cite{hsiao2004prediction}, which emits bubbles into the liquid according to local liquid/gas interface flow conditions based on the sub-grid air entrainment modeling proposed by Ma et al. \cite{ma2011comprehensive}, and tracks all bubbles in the liquid flow using equations of motion and bubble dynamics. Further breakup of the dispersed bubbles into smaller ones is modeled by incorporating the bubble breakup model developed by Martinez-Bazan et al. \cite{martinez1999breakup}. The software was parallelized by using a hybrid MPI-OpenMP scheme.

The video describes the impinging of a water jet with a diameter of 4 cm at a velocity of 4 m/s, the entrainment and breakup of the air pockets accompanied by strong production and interaction with vorticity structures, and the subsequent bubbly flows. This video has been submitted to the \emph{Gallery of Fluid Motion} 2011, which is an annual showcase of fluid dynamics videos. The described video is available at the following URLs:

\vspace*{0.2in}
\indent  $\bullet$ ~~\href{hiRes_airEntrainmentModeling.mpg}{Video 1} -- High resolution \\
\indent  $\bullet$ ~~\href{LowRes_airEntrainmentModeling.mpg}{Video 2} -- Low resolution 
\vspace*{0.2in}

This work was supported under STTR Phases I and II program, contract No. N66604-08-C-0798, funded by ONR and monitored by Dr. William Keith in NUWC, and conducted in cooperation with Pen State University and the University of Wisconsin.  


\begin{thebibliography}{10}
\expandafter\ifx\csname url\endcsname\relax
  \def\url#1{\texttt{#1}}\fi
\expandafter\ifx\csname urlprefix\endcsname\relax\def\urlprefix{URL }\fi

\bibitem{hsiao2004prediction}
C.T. Hsiao and G.~Chahine.
\newblock Prediction of tip vortex cavitation inception using coupled spherical
  and nonspherical bubble models and navier--stokes computations.
\newblock \emph{Journal of marine science and technology}, 8\penalty0
  (3):\penalty0 99--108, 2004.

\bibitem{ma2011comprehensive}
J.~Ma, A.A. Oberai, D.A. Drew, R.T. Lahey, and M.C. Hyman.
\newblock A comprehensive sub-grid air entrainment model for rans modeling of
  free-surface bubbly flows.
\newblock \emph{The Journal of Computational Multiphase Flows}, 3\penalty0
  (1):\penalty0 41--56, 2011.

\bibitem{martinez1999breakup}
C.~Mart{\'\i}nez~Baz{\'a}n, J.L. Monta{\~n}{\'e}s~Garc{\'\i}a, and J.C.
  Lasheras.
\newblock On the breakup of an air bubble injected into a fully developed
  turbulent flow. part 2. size pdf of the resulting daughter bubbles.
\newblock \emph{Journal Fluid Mechanics}, 401:\penalty0 183--207, 1999.

\end{thebibliography}
\end{document}